\UseRawInputEncoding
\documentclass[journal,onecolumn]{IEEEtran}
\usepackage{amsmath}                
\usepackage{amsthm}                 
\usepackage{amssymb}
\usepackage{thmtools}
\usepackage{kpfonts}
\usepackage[geometry]{ifsym}
\usepackage{dsfont}
\usepackage{multirow}
\usepackage[mathscr]{euscript}
\usepackage{graphicx}
\usepackage{color}
\usepackage{paralist}
\usepackage{framed}
\usepackage{float}
\usepackage{longtable}
\usepackage{cite}
\usepackage[inline]{enumitem}

\usepackage[colorlinks=true, citecolor=blue, linkcolor=blue]{hyperref}       
\usepackage[font=itshape]{quoting}
\usepackage{lscape}
\usepackage{makecell}
\usepackage[section]{placeins}
\usepackage{subcaption}

\usepackage{listings}
\usepackage{comment}
\usepackage{framed} 
\usepackage{nomencl} 
\makenomenclature

\declaretheoremstyle[%
  headfont=\bfseries,%
  headpunct={:},%
  notefont=\normalfont\bfseries,%
  notebraces={--~}{},
    qed=$\blacksquare$,
]{definitionstyle}
\theoremstyle{definition}
\declaretheorem[style=definitionstyle,name=Definition]{defn}

\theoremstyle{definition}

\theoremstyle{plain}
\theoremstyle{remark}

\usepackage{authblk}
\begin{document}
%
\title{A Hetero-functional Graph Theory Perspective of Engineering Management of Mega-Projects}
\author[1]{Amirreza Hosseini \thanks{Corrresponding Author: shossein1@stevens.edu}}
\author[1]{Amro M. Farid}
\affil[1]{Department of Systems Engineering, Charles V. Schaefer, Jr. School of Engineering and Science, Stevens Institute of Technology, Hoboken, NJ, USA}

\date{May, 2025}
\maketitle

\begin{abstract}
Megaprojects are large-scale, complex, and one-off engineering endeavors that require significant investments from a public or private sector. Such projects generally cost more than a billion dollars, take many years to develop and construct, involve stakeholders both in the public and private sectors, and impact millions of people.  Most of the extant megaproject research is concerned with understanding why the engineering management of megaprojects fails so frequently and which dimensions make them so difficult to manage, including size, uncertainty, complexity, urgency, and institutional structure \cite{denicol:2020:00}.   Recently, the literature on mega-projects has advocated for a convergence of the engineering management and production system management literature.   
To that end, this paper proposes the use of Model-Based System Engineering (MBSE) and Hetero-Functional Graph Theory (HFGT), where the latter, quite interestingly, finds its origins in the mass-customized production system literature.  More specifically, HFGT was developed so that the physical and informatic parts of production system planning, operations, and decision-making are readily reconfigured to support production customization at scale.  As the literature on megaprojects is rapidly evolving with a significant amount of divergence between authors, this report builds upon the recent and extensive megaproject literature review provided by Denicol et. al. \cite{denicol:2020:00}.  The paper concludes that MBSE and HFGT provide a means for addressing many of the concluding recommendations provided by Denicol et. al.  MBSE and HFGT not only align with current research on megaprojects but also push the boundaries of how the engineering management of megaprojects can gain a unified theoretical foundation.    
\end{abstract}



\section{Introduction: Motivation}
Megaprojects are large-scale, complex, and one-off investments in a variety of public and private sectors. They generally cost more than a billion dollars, take many years to develop and construct, involve stakeholders both in the public and private sectors, and impact millions of people. Examples of megaprojects include: the Suez Canal, the Montreal Summer Olympics, and the Sydney Opera House \cite{flyvbjerg:2014:00}.  Based on the Oxford Handbook of Mega-Projects, they are designed ambitiously to change the structure of society in spite of the fact that some small projects in the same domain have more well-established precedents\cite{flyvbjerg:2017:00}. Megaprojects, however, are not just a magnified version of a smaller project in the same domain, they are often a completely different type of project in terms of the level of aspiration, stakeholder involvement, lead-time, complexity, and impact\cite{flyvbjerg:2017:00}. Consequently, they also demand a greater level of leadership.  Flyvberjg estimates the global expenditure on megaprojects at US6-9 trillion dollars annually\cite{flyvbjerg:2014:00}.
Consequently, the management of these complex projects has garnered attention for more intensive research, along with a growing need to understand the specific governance challenges they present \cite{AHERN:2014:1371}. Flyvbjerg, in his book \cite{flyvbjerg:2003:03}, refers to these multibillion-dollar megaprojects as a new 'political and physical animal' encountered across the world. Several studies, including \cite{han:2024:00}, indicate that the number of papers featuring keywords related to the megaproject management literature has more than doubled during the 20-year period from 2004 to 2023, compared to the period from 1984 to 2003. Additionally, the growing interconnectivity among nations and the pressing need for infrastructure development in developing countries to match the standards of developed nations have led to an increasing number of megaprojects. These projects often involve international collaboration and present more complex risks than traditional megaprojects, necessitating advanced risk management strategies to overcome both foreseen and unforeseen risks such as disasters and distruptions \cite{kardes:2013:00, flyvbjerg:2009:04, yaghmaei:2022:00}. 

Most of the extant megaproject research is concerned with understanding why megaprojects fail so frequently and which dimensions make megaprojects so difficult to manage including size, uncertainty, complexity, urgency, and institutional structure \cite{denicol:2020:00}.  A review of the existing literature reveals that early research on megaprojects predominantly focused on construction and infrastructure projects, emphasizing traditional project management dimensions such as cost, time, and benefits. However, over the past two decades, the scope of megaproject research has gradually broadened to include complex R\&D (research and development) projects and public-private partnerships (PPPs). Our review highlights that recent studies increasingly focus on keywords such as performance, complexity, risk, innovation, and decision-making \cite{han:2024:00}.
This complexity and difficulty extend from the organizational environment of these megaprojects to risks within the supply chain and process modeling \cite{wickramatillake:2007:00,locatelli:2014:00,stefano:2023:00}. However, the degree of complexity is not uniform across all projects \cite{giezen:2012:00}.
The Iron Law of Megaproject Management states that 99.5\% of megaprojects fail to stay within budget, finish on time, and deliver anticipated benefits.  Flyvbjerg recognizes that if approximately one out of ten megaprojects remains within budget, one out of ten stays on schedule, and one out of ten delivers the promised benefits, then approximately one in one thousand projects is an ``on-target" success that meets all three goals\cite{flyvbjerg:2014:00}.  Even if the success rate were to double to 2 out of 10 for each criterion, then only 8 out of 1000 would be successful. \textcolor{black}{Therefore, the majority of megaprojects experience cost overruns, delayed completion, and technical issues \cite{koppenjan:2011:00}.}Consequently, the Iron Law of Megaproject Management imposes a strong call to action \cite{flyvbjerg:2014:00}. \textcolor{black}{ Additionally, certain megaprojects, such as the Boston Big Dig and the Los Angeles Subway, have become notorious for their extensive cost overruns \cite{lehrer:2008:00, erfani:2024:00}. Cost overruns and benefit shortfalls are the most common phenomena in megaprojects \cite{flyvbjerg:2009:04,flyvbjerg:2018:02,assaf:2006:00,stefano:2023:00}, and have been rigorously analyzed in the literature \cite{flyvbjerg:2003:07}. In some instances, there is evidence of intentional misrepresentation and misinformation contributing to these overruns \cite{flyvbjerg:2002:06}. Nevertheless, a few projects have been completed on time and within budget, only to face immediate operational failures. This paradox is aptly captured by the old adage: ``The surgery was a success, but the patient died." Thus, it is crucial to distinguish between executing a project successfully and achieving the intended outcomes \cite{locatelli:2014:00,zhang:2024:00}.  }    Moreover, Denicol et. al find that, at present, there is no single concept or framework that can account for the multiple causes of and cures for poor megaproject performance\cite{denicol:2020:00}.  

In addition to the previously mentioned causes of megaproject failure, several studies highlight the role of organizational culture and strategies to strengthen it for more effective collaboration \cite{vanmarrewijk:2008:00, vanmarrewijk:2006:00, martin:2001:00}. Some researchers argue that the diversity of organizational cultures within megaprojects presents an alternative perspective to Flyvbjerg's view \cite{flyvbjerg:2003:03}, which attributes megaproject motivations primarily to vested interests \cite{vanmarrewijk:2008:00}. Martin’s approach, which focuses on ambiguity, power dynamics, and situational factors in organizational culture, provides valuable insights for addressing the daily challenges of megaprojects \cite{martin:2001:00}.
The concept of megaproject culture is characterized by inherent ambiguity, with unclear boundaries and a duality between the tangible elements and the stakeholders striving to bring the project to fruition \cite{engwall:1998:00}. This complexity has driven research interest in fostering cooperation and mitigating opportunistic behavior—actions where firms prioritize their self-interest at the expense of other stakeholders \cite{williamson:2007:00}. To address these issues, a shift away from traditional contracting approaches is advocated. Traditional methods, which relied on fixed parameters and measurable risks, are effective for known risks but fail to address "unknown-unknowns" \cite{dequech:2011:00}.
Under traditional contracts, principals sought to transfer risks to agents, while agents aimed to secure reasonable returns for their efforts, often resulting in opportunistic behaviors. To mitigate such risks, alliance contracting—emphasizing trust, governance, and cultural alignment—has been introduced. This approach complements traditional risk management strategies and addresses risks that contribute to cost overruns, delays, and the failure to deliver expected benefits \cite{galvin:2021:00}.

Recently, the literature on mega-projects has advocated for a convergence of the engineering management and production system management literature\cite{ma:2025:00}.   To that end, This paper proposes the use of Model Based System Engineering (MBSE) and Hetero-Functional Graph Theory (HFGT) to address the pressing gaps in the engineering management of mega-projects literature.  Quite interesting, HFGT finds its origins in the mass-customized production system literature and was developed so that the physical and informatic parts of production system planning, operations, and decision-making are readily reconfigured to support production customization at scale.  This critical review of the literature argues that MBSE and HFGT can model the complexity of megaprojects and support analyses that can bring their costs and schedule back under control.  As the megaproject management literature is rapidly evolving with a significant amount of divergence between authors, this report organizes itself around the recent and extensive megaproject management review provided by Denicol et. al. \cite{denicol:2020:00}.  More specifically, Denicol et. al concludes with four self-reinforcing research directions:  
\begin{enumerate*}
\item design the megaproject management system architecture, 
\item bridge the theoretical gap between engineering management of megaprojects and production system management, 
\item evaluate the potential for collaborative decision-making, and 
\item address the significant supply chain challenges associated with megaproject management. 
\end{enumerate*}
Consequently, this report aligns its critical perspective on the literature with these four research directions as a proxy for engineering management of megaproject literature as a whole.  Our research focuses on advancing the application of MBSE and HFGT to engineering management of megaproject, particularly in response to the challenges highlighted by Denicol et. al.  Furthermore, to facilitate the argument, the management of a mega-project is classified as a convergent system-of-systems where the mega-project requires multiple unlike systems to be integrated into a single system; and done so in a manner that utilizes a common (i.e. convergent) language.  

The remainder of this paper proceeds as follows.  Section \ref{background} provides a background to (formal) graph theory, multi-layer networks, and their limitations.  It also exposits the HFGT meta-architecture as a means of overcoming these limitations.  Section \ref{apply hfgt} then relates the HFGT meta-architecture to the complexities found in megaproject management.  Section \ref{megaproject_themes} then elaborates HFGT's potential to address six key themes found in the megaproject management literature.  Finally, section \ref{conclusion} concludes the work.

\section{Background:  Model-Based Systems Engineering (MBSE) \& Hetero-functional Graph Theory (HFGT)} \label{background}

This paper's proposed use of MBSE and HFGT is based upon the recognition that many project management techniques are ultimately rooted in graph theory and multi-layer networks.  Consequently, this section introduces graph theory in Sec. \ref{graph_limit} and multi-layer networks in Sec. \ref{network_limit} and some of the limitations that may impede their utilization in the context of mega-project management.  Sec. \ref{hfgt_architecture} then discusses the HFGT meta-architecture as a means of overcoming some of these limitations.  

\subsection{Graph Theory and its Limitations}\label{graph_limit}
Graph theory has presented a useful abstraction across many applications because it is predicated on the construction of a graph.  
\begin{defn}[Graph \label{Defn:Graph}]\cite{Steen:2010:00}
G = $\{V, E\}$, consists of a collection of nodes V and a collection of edges E. Each edge $e \in E$ is said to join two nodes which are called its endpoints. If e joins $v_1,v_2$ in V, we write e = $(v_1,v_2)$. Nodes $v_1$ and $v_2$, in this case, are said to be adjacent. Edge e is said to be incident with nodes $v_1$ and $v_2$ respectively.
\end{defn}

\begin{figure}
    \centering
    \includegraphics[width=0.5\linewidth]{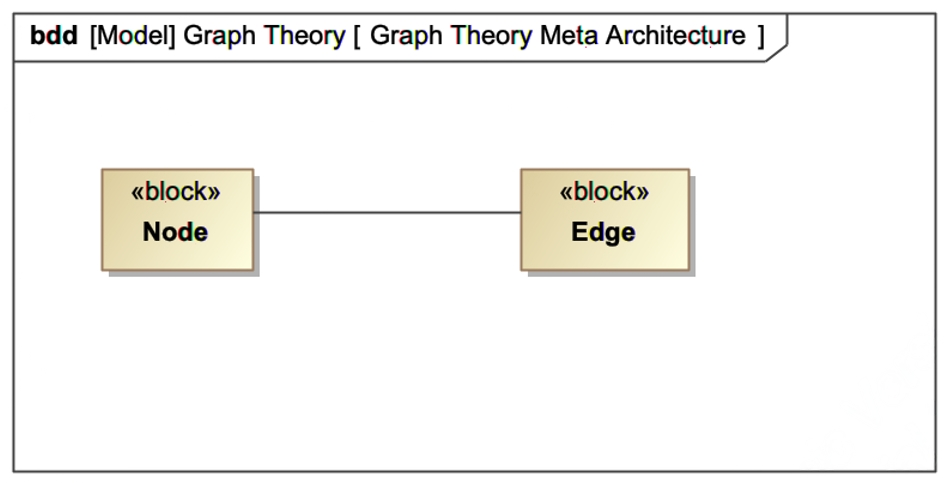}
    \caption{Block Definition Diagram of the Graph Theory Meta Architecture}
    \label{fig:graphBDD}
\end{figure}


While the above definition recognizes that all systems are composed of interconnected constituent elements, it has limitations in the systems engineering of convergent systems-of-systems.  From a MBSE perspective, Defn. \ref{Defn:Graph} implies a graph theory ``meta-architecture" as depicted in the block definition diagram in Fig. \ref{fig:graphBDD}
A block definition diagram of the graph theory meta-architecture is presented in Fig.\ref{fig:graphBDD}.  Here, a meta-architecture is an architecture that describes multiple systems in a domain-agnostic language/ontology without reference to the multiplicity caused by instantiation.  Consequently, a transportation system might have many intersections and roads, but the intersections are still nodes, and roads are still edges.  Note that graph theory, as it is traditionally and most commonly applied, focuses primarily on an abstracted model of a system’s form; neglecting an explicit description of the system’s function. The system function, what a system does, in terms of verbal phrases, has been entirely omitted from the explicit statement of the formal graph and any understanding of the system’s function is implicit.  For example, the same transportation system may be used to allow vehicles to move along roads from one intersection to another, but this transportation function is implied by the graph rather than explicitly stated.  Consequently, the graph theory meta-architecture is quite limiting and it is less than clear how formal graph theory may be applied to convergent systems-of-systems that are fundamentally transformative with multiple operands. As convergent systems-of-systems are \textit{hetero-functional}, formal graph theory may impede rigorous approaches where multiple and unlike megaproject management functions and operands can be modeled.  

\subsection{Limitations of Multi-Layer Networks}\label{network_limit}
To overcome many of the analytical limitations of (formal) graph theory, the network science literature has advanced the concept of ``multi-layer networks" where two or more network “layers" interact with each other.
\begin{defn}[Multi-Layer Networks \label{Defn:Multi-Layer Networks}]\cite{Kivela:2014:00}
Much like a conventional graph, a multi-layer network $G_m = \{V_m , E_m \}$ is formally defined as a tuple of nodes $V_m$ and edges $E_m$. Such a multi-layer network is organized into an integer n number of layers $L_1 ... L_n$. Here, a given layer $L_\alpha = \{V_\alpha,E_\alpha\}$ is understood as a graph where the nodes $V_\alpha$ and edges $E_\alpha$ have at least one semantic aspect, feature, or operand in common (e.g. electricity, water, people, etc.) 
\end{defn}

\begin{figure}[ht]
    \centering
    \begin{subfigure}{0.48\textwidth}
        \centering
        \includegraphics[width=\linewidth]{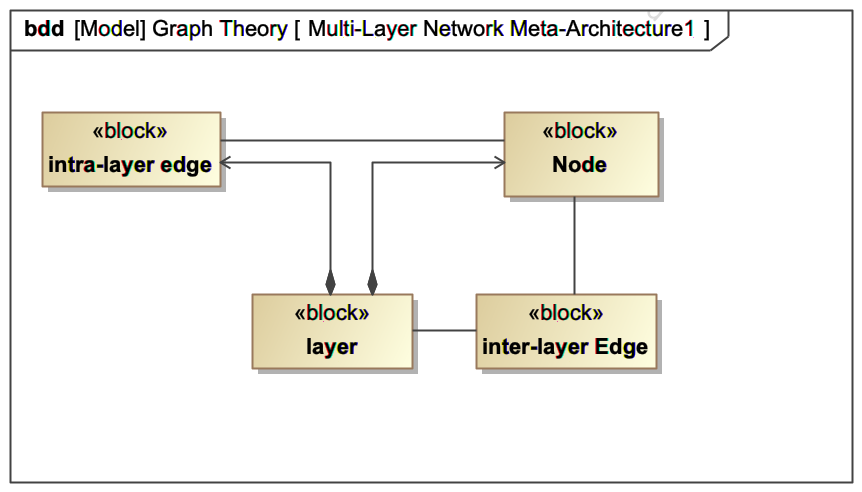}
        \caption{Block Definition Diagram of a Multi-Layer Network Meta Architecture}
        \label{fig:multilayer1}
    \end{subfigure}
    \hfill
    \begin{subfigure}{0.48\textwidth}
        \centering
        \includegraphics[width=\linewidth]{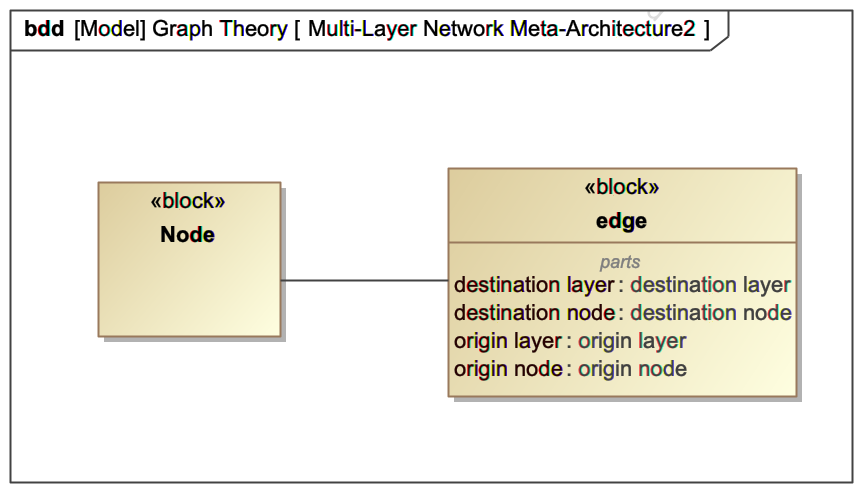}
        \caption{An Alternative Block Definition Diagram of a Multi-Layer Network Meta Architecture}
        \label{fig:multilayer2}
    \end{subfigure}
    \caption{Multi-layer Network Meta Architecture Representations}
    \label{fig:multilayer_all}
\end{figure}

From an MBSE perspective, Defn. \ref{Defn:Multi-Layer Networks} multi-layer network implies the meta-architecture depicted in Fig. \ref{fig:multilayer1}.  The node block is retained, and the edge block becomes an intra-layer edge block.  Both of these become part of a layer as a new block. Finally, an inter-layer edge is added with associations to two nodes and two layers.  Alternatively, a multi-layer network can be conceptualized using the meta-architecture in Fig. \ref{fig:multilayer2}.  In this case, a multi-layer network retains the node and edge blocks from the graph theory meta-architecture in Fig. \ref{fig:graphBDD}, but makes sure to label each node with its associated layer (attribute) and each edge with its origin and destination node and layer.

\begin{figure*}[h]
\centering
\includegraphics[width=0.7\linewidth]{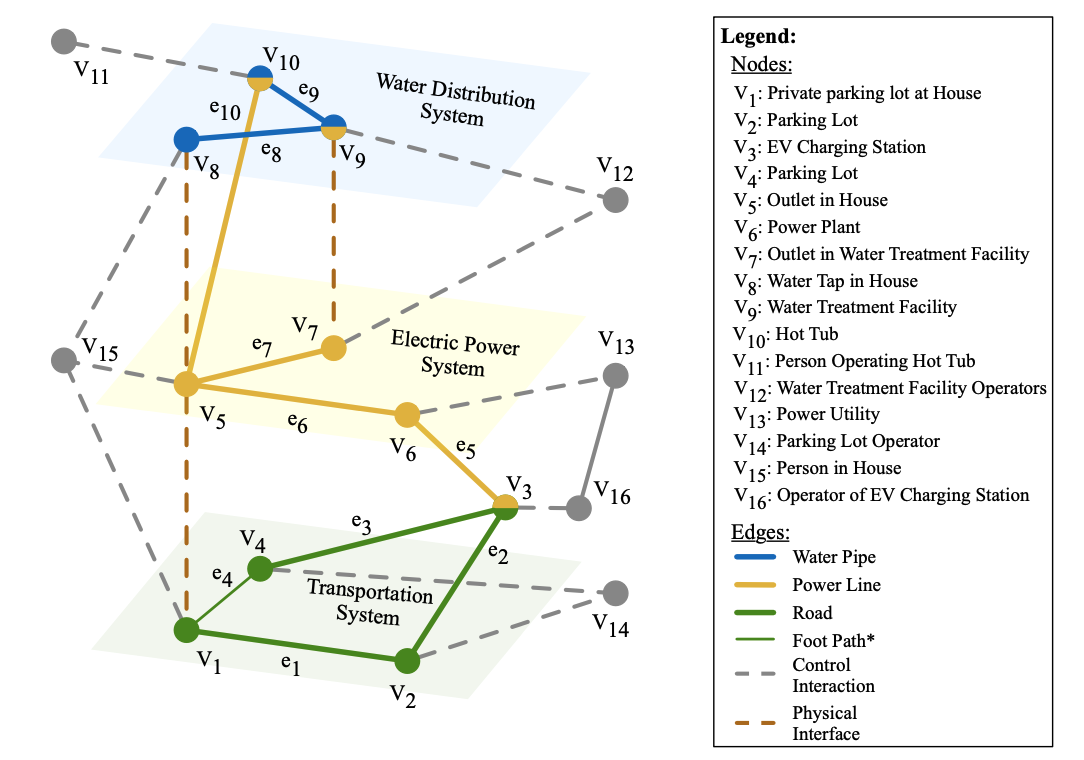}
\caption{A Hypothetical Four-Layer Network including transportation, electric power, water distribution, and decision-making elements that exhibit all eight of the modeling constraints found in multi-layer networks.\cite{Schoonenberg:2019:ISC-BK04}}
\label{fig:exmulti-layer network}
\end{figure*}

Despite the expanded meta-architecture of multi-layer networks, Kivela et. al, in their comprehensive review, showed that \textbf{all} multi-layer network models have from at least one of the following limitations \cite{Kivela:2014:00}:
\begin{enumerate}
    \item Alignment of Nodes between layers
    \item Disjointment between layers
    \item Equal number of nodes for all layers
    \item Exclusively vertical couplings between layers
    \item Equal couplings between layers
    \item Node counterparts are coupled between all layers
    \item Limited number of modeled layers
    \item Limited number of aspects for each layer
\end{enumerate}
To demonstrate the consequences of these modeling limitations, the HFGT textbook conceived the hypothetical system shown in Fig. \ref{fig:exmulti-layer network} as an example that cannot, at present, be modeled with multi-layer networks.  In contrast, a complete HFGT analysis of this hypothetical test case was demonstrated in the aforementioned text\cite{Schoonenberg:2019:ISC-BK04}.  To follow up this result, the tensor formulation of hetero-functional graph theory proved that multi-layer networks are neither ontologically lucid nor complete\cite{Farid:2022:ISC-J51}.  As the management of mega-projects is likely to exhibit at least one of the characteristics described above, multi-layer networks are likely to present modeling limitations that result in analytical impasses.  

\subsection{The HFGT Meta-Architecture}\label{hfgt_architecture}
Sections \ref{graph_limit} and \ref{network_limit} revealed that (formal) graph theory and multi-layer networks exhibit ontological limitations that impose analytical limitations in the context of mega-project management.  Ultimately, one of the main sources of complexity in mega-projects is not just their size but also their heterogeneity in structure and function.  In contrast, HFGT has a rich meta-architecture that has the potential to address this heterogeneity.  While a complete treatment of HFGT is not feasible here, a brief exposition elucidates its value to megaproject management.  

\begin{figure*}[h]
\centering
\includegraphics[width=1.0\linewidth]{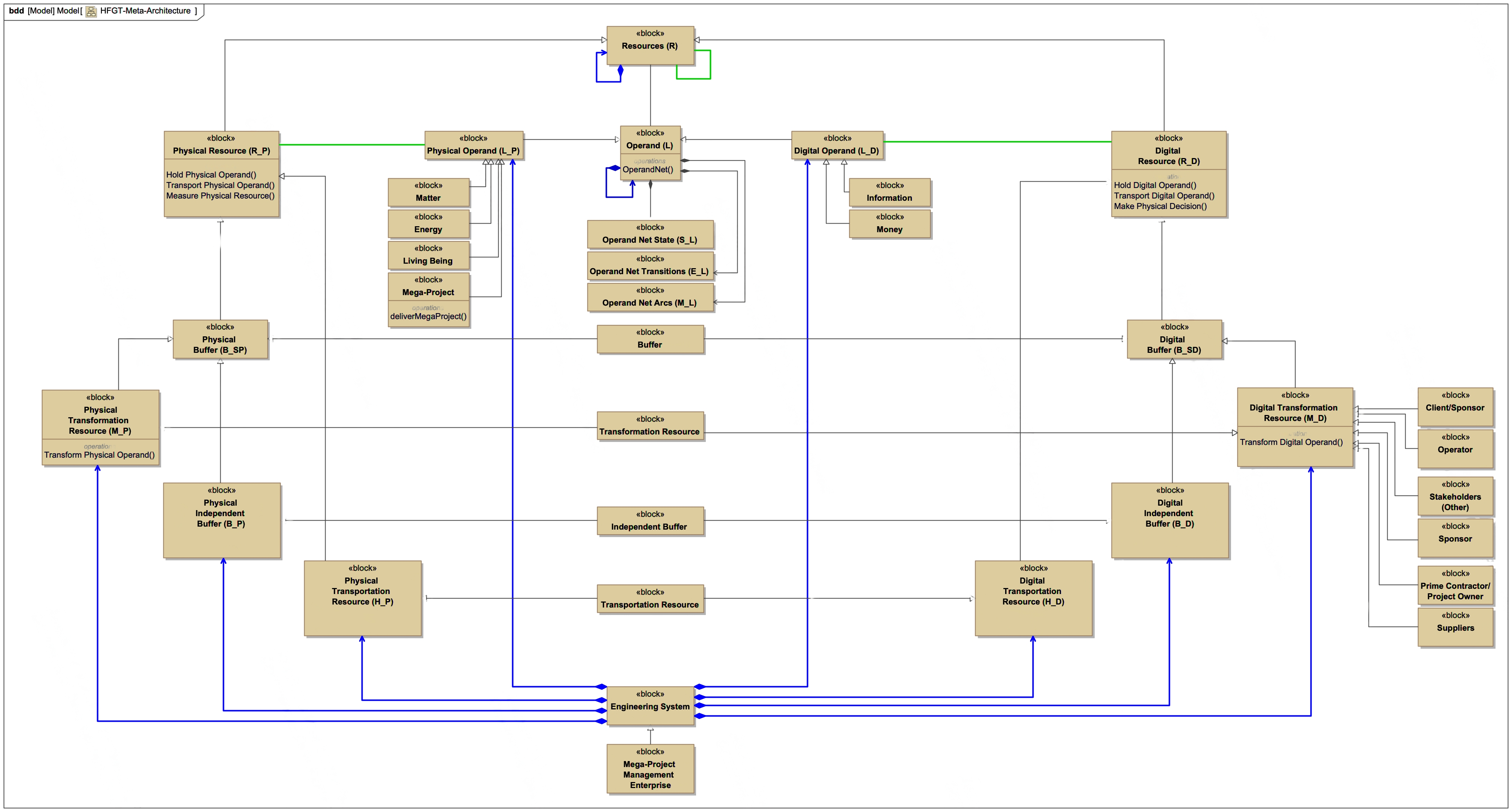}
\caption{Block Definition Diagram of the HFGT Meta Architecture}
\label{fig:HFGT meta}
\end{figure*}

Rather than using nodes and edges as its starting point, hetero-functional graph theory recognizes that humans' ability to describe (engineering) systems is rooted in natural language and more specifically subject + verb + object sentences.  Consequently, the HFGT meta-architecture shown in Fig. \ref{fig:HFGT meta} states that engineering systems (e.g. mega-project management enterprises) are composed of (digital and physical) operands as objects and (physical and digital ) resources as subjects.  Naturally, these physical and digital resources also carry out processes as verb + operand predicates.  Ultimately, the combination of a resource and a process creates a capability that serves as a node in a hetero-functional graph and the logical sequence of two sentence-capabilities becomes an edge.  

Building upon the central theme of subject-verb-object sentences, the HFGT meta-architecture in Fig. \ref{fig:HFGT meta} recognizes that there are fundamental differences between the types of operands, resources, and processes.  These differences -- when recognized within the ontology of HFGT -- directly support analyses of complex engineering systems like mega-projects.  Physical operands are classified into matter, energy, and living beings because they exhibit distinguishing features that are lost if the categories are merged. Consequently, HFGT treats a megaproject as an aggregated physical operand (composed of matter, energy, and living beings) that a mega-project management enterprise (as an engineering system) must deliver.  This mega-project (as a physical operand) has a state that is initialized at its start, evolves through transitions connected by arcs as it proceeds, and ultimately concludes.  The mega-project management enterprise is also composed of transformation resources, independent buffers, and transportation resources (as physical resources) that realize transformation, holding, and transportation processes on the mega-project and its constituent matter, energy, and living being operands.  The distinction between transformation and transportation stems from the preference toward value-adding (vs non-value-adding) activities in manufacturing.  In the meantime, the spatial distribution of complex engineering systems (like engineering management of megaproject) necessitates transportation processes.  The mega-project management enterprise is also composed of digital-operands that represent information and money.  These underscore the digital and financial considerations inherent to complex engineering systems.  Finally, a mega-project management enterprise is composed of a plethora of digital resources including clients/sponsors, prime contractors, suppliers, operators, and other stakeholders.  These digital resources have explicit roles that include measurement processes (that pull information from physical resources and operands), decision processes (that send information to physical resources and operands, digital-transformation processes (that create/destroy digital-operands e.g. creation of money/information), and digital-transportation processes (e.g. money transfer and information communication).  In all, HFGT recognizes the fundamental differences between types of operands, resources, and processes while still retaining the underlying simplicity of subject + verb + object sentences.  

In addition, HFGT has demonstrated its versatility and broad applicability across many scientific and engineering disciplines; underscoring its potential in convergence science.  HFGT's ability to model heterogeneous networks of arbitrary topology and connect them in flexible ways surpasses the capabilities of traditional multi-layer networks. This adaptability makes HFGT valuable to a wide range of professionals. For natural and engineering scientists, HFGT provides a framework to reconstitute conservation laws within complex systems. Decision scientists can use HFGT to support diverse decision-making structures, whether centralized, decentralized, or distributed. Systems engineers find a natural translation from SysML into quantitative models with HFGT, while operations researchers benefit from its generalization of minimum cost flow principles. Furthermore, social scientists can bridge the gap between qualitative insights and quantitative models using HFGT, and applied mathematicians can extend their work with the theory's solid foundation in graph theory and tensor analysis. Collectively, these attributes demonstrate HFGT’s significant potential to unify disparate fields within convergence science, fostering interdisciplinary collaboration and innovation.

Finally, HFGT has made foundational contributions to the structural analysis of complex systems. HFGT's versatility allows it to be applied to complex systems of systems with arbitrary topologies, making it highly adaptable to various urban and engineering contexts. The theory’s extensibility enables it to incorporate numerous physical elements and functions, ensuring comprehensive coverage of the system under analysis. HFGT is inherently designed to manage directed graphs, which are crucial for representing the directional flow of resources and information in complex systems. Moreover, HFGT is fundamentally cyber-physical, integrating both physical and informatic aspects of system architecture. Ultimately, HFGT posits that beneath the specific applications lies an underlying meta-architecture that can be applied generically across different domains, offering a unified and convergent framework for understanding and optimizing a wide range of complex engineering systems.

\section{Applying the HFGT Meta-Architecture to Megaproject Management}\label{apply hfgt}
In light of the HFGT meta-architecture introduced in Sec. \ref{hfgt_architecture}, this section now seeks to relate it to the complexities of megaproject management that Denicol et. al articulate in the introduction of their literature review.  The HFGT meta-architecture provides a systematic and structured framework for bridging the gap between theoretical/quantitative models and the practical challenges in managing large-scale projects.  The following discussion explores how HFGT can address key aspects of megaproject management including the coordination of diverse stakeholders, the decomposition of complex projects into manageable components, and the integration of decision-making processes across various levels of project execution. 
\begin{itemize}
\item Megaprojects are inherently established as temporary organizations, led by various entities such as client teams, prime contractors, or coalitions of multiple parties, which may include owners, sponsors, and other stakeholders\cite{zani:2024:00}. These entities collaborate on a shared activity within an uncertain environment\cite{zani:2024:01}. HFGT offers a systematic approach to modeling these complex, temporary structures by decomposing them into key components.  First, HFGT conceptualizes the megaproject itself as an operand that may have an uncertain state and evolution.  Furthermore, the megaproject management enterprise creates the organizational structures for the conduct of the megaproject itself.  Its digital resources include stakeholders, prime contractors, temporary alliances, joint ventures, and coalitions of multiple parties. Such a model transparently identifies the complex interrelationships between all components of the megaproject and their inherent uncertainties.
\item Megaprojects complexity  often requires them to be decomposed into smaller, interrelated projects, which are then organized as part of a larger program\cite{kian:2017:00, nyarirangwe:2019:00}. HFGT addresses this complexity by treating the megaproject as an operand that may be decomposed into additional operands that represent projects in and of themselves.  Because operands can be systematically decomposed into smaller parts, it is possible to implement a structured approach to managing the detailed relationships and dependencies between smaller projects and the larger megaproject.  In addition to a mathematical description, the visualization of the mega-project architecture facilitates the seamless integration of the component projects and augments the potential for effective coordination and management.  
\item Megaprojects also require robust organizational structures to coordinate the efforts of multiple entities, such as the client, prime contractor, and suppliers\cite{olawale:2020:00}. HFGT effectively models these complex organizational structures within a mega-project management enterprise composed of digital resources that include these entities.  In HFGT, these digital resources have agency over their respective physical resources (i.e.technical assets and human resources).  This creates a mega-project management enterprise architecture where digital resources and their processes are integrated with physical resources and their processes and these are in turn applied to the execution of the megaproject itself.  This architectural description can enhance transparent oversight over and alignment between all aspects of the project including the coordination between subgroups and the management of physical, financial, and informatic interfaces.  Furthermore, HFGT's ability to quantitatively model these interactions is vital to optimizing time, cost, and quality goals for all stakeholders involved.  
\item Mega-project management suffers from distinct theoretical foundations and analytical frameworks that collectively lead to piecemeal approaches and solutions\cite{cantarelli:2023:00, pitsis:2018:00}.  In contrast, HFGT converges all aspects of the mega-project management enterprise within its meta-architecture as a unifying ontological and architectural framework.  More specifically, digital resources and their processes can provide a cohesive understanding of decision-making, leadership, teamwork, and team integration within a unified system.  Returning to the meta-architecture in Fig.\ref{fig:HFGT meta}, the decomposition arrows (in black) and association links (in green) identify these relationships explicitly.  While traditional leadership, in a socio-psychological sense, is not directly modeled, HFGT captures the hierarchical structure and peer-to-peer interactions between decision-makers.  Therefore, it can offer a comprehensive framework for analyzing the multifaceted nature of megaproject management enterprises.  Furthermore, HFGT asserts that such a meta-architecture can serve as a unifying theoretical foundation for the complex, integrated, and quantitative analyses required in megaproject management. Such an integrated approach can serve to reconcile disparate and partial theoretical approaches and ultimately offer a cohesive strategy for understanding and managing large-scale, multifaceted projects.
\end{itemize}

In all, Denicol et al. emphasize the need for new theory-building research that adopts a systemic view founded in production systems and integrates the various aspects of megaproject management performance\cite{denicol:2020:00}. HFGT aligns with this recommendation by offering a unifying theoretical foundation and quantitative framework for the analysis of megaprojects.  Furthermore, it is important to recognize that the earliest works on HFGT address mass-customized production systems that bear many of the same features found in mega-projects.  Denicol further observes that the terminology within the field of megaproject management lacks convergence.  Different authors use different conceptualizations and different terms to describe the same phenomena\cite{wangg:2023:00, ashkanani:2023:00}.  In contrast, HFGT imposes a consistent and convergent terminology throughout its meta-architecture.  Operand, resource, and process definitions are clearly stated and their classifications are transparently made.  This meta-architecture has already proven to be effective across a range of domains, including mass-customized production systems\cite{Farid:2007:IEM-TP00, Farid:2013:EWN-C15,Farid:2015:ISC-J19, Farid:2014:ISC-C37, Farid:2014:ISC-C38, Farid:2008:IEM-J04}, transportation systems\cite{Viswanath:2014:00, Baca:2013:ETS-C11,Baca:2013:ETS-C23}, energy systems\cite{Farid:2015:SPG-J17,Farid:2014:SPG-C42,Hingorani:2000:00}, water systems\cite{Lubega:2014:00} and healthcare systems\cite{Khayal:2015:ISC-J20}.Most recently, HFGT's potential for convergence has been recognized with a \$3.6M NSF Growing Convergence Research project\cite{NSF:2024:00}.  
This extensive evidence across a wide variety of disparate application domains suggests that HFGT's approach to imposing a unified terminology and framework is well-suited to the domain of megaprojects. 

\section{Addressing Megaproject Management Themes}\label{megaproject_themes}
In addition to their introductory exposition on engineering management of megaproject, Denicol et al. analyzed 86 articles to identify six key themes that impact megaproject performance: 
\begin{enumerate*}
\item decision-making behavior
\item strategy and governance
\item risk and uncertainty
\item leadership and capable teams
\item stakeholder engagement
\item supply chain integration 
\end{enumerate*}. 
This section elaborates HFGT's potential to address these six key themes.  

\subsection{Decision-Making Behavior}
In general, HFGT addresses decision-making through its digital resources, the digital processes allocated to them, and the serial and parallel arrangement of these processes.  Denicol et al. and some other studies such as \cite{flyvbjerg:2003:03, bruzelius:2002:00, chadee:2021:00, montrimas:2021:00,Galli:2020:00, love:2018:00, lehtonen:2014:00, nocera:2018:00, boateng:2015:00} further elaborate three critical issues under the decision-making behavior theme:
\begin{enumerate*}
\item optimism bias,
\item strategic misrepresentation,
\item escalating commitment.
\end{enumerate*}
To mitigate optimism bias, HFGT has produced stochastic discrete-event simulation models that offer a more realistic depiction of system execution, potentially reducing overly optimistic expectations, and allowing for a more accurate assessment of potential risks\cite{Khayal:2021:ISC-J48}.  Additionally, such stochastic discrete-event simulation models can quantify the negative impacts of optimism bias on megaproject management lead times, delivery, and resilience.  Strategic misrepresentation is a more complex socio-cultural issue where decision-makers may mislead or be incentivized to mislead stakeholders.  An HFGT model is unlikely to resolve such a problem because all models are only as good as the data they are built upon.  Nevertheless, the development of a consistent and unified framework is much more likely to transparently reveal inconsistent modeling assumptions and data; thereby creating a moderate deterrent against strategic misrepresentation.  Finally, HFGT addresses escalating commitments where decision-makers persist in a failing direction of project execution and ignore alternative pathways to project completion. Because HFGT is ultimately a type of graph theory, it identifies lower-risk options and execution pathways as alternative paths through a hetero-functional graph\cite{Farid:2015:ISC-J19} and thereby facilitates their selection.   By applying these analyses, HFGT enhances the rationality of decision-making processes, mitigates common behavioral pitfalls in megaproject management, and supports more effective and informed decision-making throughout the project.

\subsection{Strategy, Governance, and Procurement}
HFGT also addresses strategy, governance, and procurement within the structure and function of its decision-making architecture.  In this context, Denicol et al. work and some other studies \cite{wangg:2023:00, fu:2024:00, liu:2024:00, esposito:2023:00, kose:2024:01}identify three key concepts:
\begin{enumerate*}
\item the roles and responsibilities of stakeholders,
\item governance structures, 
\item delivery model strategies.
\end{enumerate*}
Perhaps one of the most confounding aspects of megaprojects is that the processes of physical execution and decision-making pertaining to the megaproject are distributed amongst a plethora of stakeholders.  There is no single entity that has the authority to compel the execution of all parts of the megaproject.   Consequently, the only hope for successful execution of the megaproject is when all stakeholders understand their respective roles and responsibilities.  Fortunately, and as mentioned previously, HFGT explicitly identifies all physical and decision-making processes in the megaproject management enterprise and allocates them to their associated physical and decision-making resources.  This allocation provides clarity to all stakeholders as to their respective roles and responsibilities.  Furthermore, the HFGT meta-architecture admits aggregations of decision-making resources and processes; thus enabling the construction of complex governance structures such as teams, committees, organizations, and consortia.  Finally, the HFGT literature provides many examples of modeling peer-to-peer relationships in supply chains in production, service, and infrastructure systems that span multiple organizations\cite{Farid:2015:ISC-J19,Thompson:2024:ISC-J55,Thompson:2023:ISC-J53,Schoonenberg:2022:ISC-J50}.

\subsection{Risk and Uncertainty}
HFGT also provides a structured approach to addressing risk and uncertainty.  In this context, several researcher  highlight three key challenges\cite{denicol:2020:00, xue:2024:00, othman:2024:00, younesi:2022:00, castelblanco:2024:00, Marcondes:2019:00, zhang:2024:01,hu:2015:01}: 
\begin{enumerate*}
\item technological novelty,
\item flexibility, and
\item complexity.
\end{enumerate*}
These challenges are critical as they influence the successful management and execution of large-scale projects. Technological novelty introduces risks associated with first-of-their-kind technologies, where balancing between the reuse of established technologies and innovative designs is essential. Flexibility refers to the system's ability to adapt to changing circumstances. Complexity involves managing the numerous parts and interrelationships within a megaproject and its external interactions. HFGT offers a structured approach to modeling and managing these challenges.  Generally speaking, risk and uncertainty are represented as stochastic quantities within the system, supported by HFGT's ability to simulate discrete events.  More specifically, and for technological novelty, HFGT distinguishes between ``known-unknowns" and ``unknown-unknowns".  For ``known-unknowns", a system's parametric model allows for detailed analysis.  Yet ``unknown-unknowns" are harder to manage as they represent design challenges that cannot be easily modeled before they exist.  In terms of flexibility, HFGT addresses design, construction, and supply chain flexibility.  More specifically, it distinguishes between the existence and availability of megaproject capabilities; recognizing that while some capabilities may not be currently utilized, it may be relatively straightforward to make them available. This distinction allows the megaproject management enterprise to respond to foreseen and unforeseen risks and systematically develop resilience as a life cycle property.  Finally, HFGT provides tools to quantify and manage complexity.  It specifically models modularity, decomposes the system into manageable modules, and controls the cascading risks across the mega-project management enterprise.  By integrating these strategies, HFGT offers a comprehensive framework that addresses the risks and uncertainties inherent in megaprojects.

\subsection{Leadership and Capable Teams}
In exploring the theme of leadership and capable teams within megaprojects, works in the literature \cite{denicol:2020:00, zaman:2024:00, yunpeng:2024:00, howell:2005:00, Marnewick:2020:00}  identify three key concepts: 
\begin{enumerate*}
    \item project leadership,
    \item competencies,
    \item capabilities.   
\end{enumerate*}
Project leadership puts emphasis on the need for dedicated champions and leaders who are committed to the project's success.  Competencies refer to the essential skills and abilities that individuals within project teams must possess.\cite{denicol:2020:00, howell:2005:00} Capabilities, on the other hand, represent the collective organizational knowledge and ability to produce specific products or services, relying on effective team collaboration\cite{denicol:2020:00,teece:2010:00}. HFGT addresses these areas by offering a structured approach to model the roles, competencies, and capabilities within an organization.  Generally speaking, leadership is captured by modeling decision-makers and capable teams as physical resources within the project.  Stakeholders are also treated as decision-makers with process-level influence.  More specifically, with respect to project leadership, it models their formal roles and clarifies how leadership functions within the megaproject enterprise management structure and goals.  Additionally, HFGT includes individual competencies as processes and these competencies functionally aggregate to form the organization's overall capabilities. By integrating competencies within a broader MBSE framework, HFGT enhances human resource management;  ensuring that the necessary skills and knowledge are utilized effectively throughout the megaproject management enterprise.  Furthermore, HFGT models the relationships between different competencies which is necessary for building team-based organizational capabilities.  Finally, HFGT has yet to address the subtle socio-cultural differences of leadership and conflict management.  To that effect, recent works on the propagation of socio-cultural and socio-psychological factors in social networks present an interesting direction for further investigation.  In all, through its structured approach, HFGT supports the alignment of leadership and team structure with overall project objectives; contributing to the successful execution of megaprojects.

\subsection{Stakeholder Engagement and Management}
In the context of stakeholder engagement and management within megaprojects, researchers identify three critical concepts\cite{denicol:2020:00,af:2024:00, eren:2019:00, afieroho:2024:00}:
\begin{enumerate*}
\item institutional context,
\item stakeholder fragmentation, and
\item community engagement.
\end{enumerate*}
The institutional context encompasses the formal organizational structures, rules, and norms that govern the megaproject \cite{denicol:2020:00, afieroho:2024:00, arda:2024:00}. Stakeholder fragmentation refers to the challenges posed by the involvement of multiple parties;  often leading to complex and sometimes conflicting interactions among stakeholders \cite{denicol:2020:00,mulholland:2020:00,Sun:2024:00}.  Community engagement ensures that local populations are considered and integrated into the megaproject’s execution \cite{denicol:2020:00, afieroho:2024:00}.  As mentioned previously, HFGT models these complexities through its decision-making architecture.  For institutional context, HFGT models formal organizational structures and rules, to clearly represent the project’s institutional framework.  However, the formality of HFGT is less able to describe informal norms and other ``unwritten rules" of an organization.  To manage stakeholder fragmentation, HFGT directly models and thus identifies the stakeholder fragmentation.  In this respect, it is particularly important to recognize when multiple stakeholders have conflicting agency over another stakeholder.  Once identified, new decision-making processes and governance mechanisms should be designed, modeled, and implemented to resolve the potential for conflicting agency and manage the diverse interests of stakeholders.  In terms of community engagement, it is important to recognize MBSE's long tradition of participatory modeling. From there, MBSE models (i.e. in SysML) can be algorithmically converted into the HFGT counterparts.  Through its structured approach and integration with MBSE, HFGT enhances the project’s ability to navigate complex stakeholder environments and provides a means for incorporating local community interests into the decision-making processes of the engineering management of megaproject enterprise.  

\subsection{Supply Chain Integration and Coordination}
Finally, HFGT supports a wide range of continuous and discretized supply chain models; offering a comprehensive framework that integrates these critical aspects of megaproject management into a unified system for analysis and optimization.  According to some works in the literature \cite{denicol:2020:00, stefano:2023:00, maylor:2018:00, riazi:2019:00, aaltonen:2018:00}, Denicol et. al, the supply chain integration and coordination theme focuses on:
\begin{enumerate*}
\item program management,
\item commercial relationships,
\item systems integration.
\end{enumerate*}
Program management involves establishing the systems, procedures, and tools necessary to monitor, control, and optimize benefits across multiple interrelated projects\cite{denicol:2020:00,wang:2019:01}. Commercial relationships refer to the formal interactions between organizations delivering projects and sub-projects \cite{denicol:2020:00, ahola:2009:00} . Systems integration is concerned with the technical and managerial capabilities required to combine various components produced by different parties into a cohesive operational asset\cite{denicol:2020:00,davies:2009:00}. HFGT provides a structured and comprehensive framework to model these critical components within a megaproject context. HFGT allows for the straightforward representation of a megaproject as an aggregation of multiple smaller projects; each managed by physical resources and digital resources.  Thus, it facilitates the clear depiction of complex interdependencies within the supply chain. By formalizing contractual relationships and supporting change order processes, HFGT enhances program management; facilitating the potential for a live digital twin representation of the megaproject's management.  Additionally, HFGT can use contract nets to model complex commercial interactions such as games or auctions and subsequently calculate performance metrics. Systems integration is inherently supported by HFGT, as it models the physical and digital operands passed between physical and decision-making resources.  This comprehensive approach ensures that all aspects of the supply chain -- from the planning and execution of individual projects to the integration of their outputs into the final deliverable  -- are effectively coordinated and controlled.  

\section{Conclusion}\label{conclusion}

This paper proposes the use of Model-Based Systems Engineering (MBSE) and Hetero-functional Graph Theory (HFGT) to address the complexities found in engineering management of megaproject.  One of the primary challenges in megaprojects is the lack of a convergent understanding of these projects as complete production systems;  encompassing planning, design, manufacturing, construction, and integration into operations. HFGT addresses this gap by providing a systematic approach that models megaprojects as convergent systems-of-systems. This method considers the interdependencies between different phases of a megaproject; enabling a more complete study of how value is created, evolved, and transferred across the megaproject management enterprise.  Model-based Systems Engineering provides a graphical means of describing these interdependencies, and HFGT translates these graphical models into their mathematical counterparts.  Furthermore, HFGT can simulate megaprojects as stochastic discrete-event systems which provides a clear, quantitative, and unbiased foundation for decision-making.  Such an analytical capability enhances quantitative objectivity in risk and uncertainty management.  Furthermore, HFGT can model the structural relationships within megaproject management to help identify single points of failure and other structural vulnerabilities.  Finally, the HFGT's attention to a system's decision-making architecture provides a basis for identifying, engaging, and managing stakeholders and their often conflicting interests.

In all, MBSE and HFGT provide a means for addressing many of the concluding recommendations provided by Denicol et. al.
\begin{itemize}
\item MBSE and HFGT not only model the megaproject engineering management enterprise, but they also provide a systematic means for its quantitative analysis.  
\item HFGT's origins in mass-customized, reconfigurable manufacturing systems facilitate the placement of engineering management of megaproject  on a firm production system management foundation.  
\item HFGT specifically addresses a diversity of collaboration/ decision-making architectures including centralized, distributed, decentralization, hierarchical, coordinated/uncoordinated, and thereby well-equipped to manage and assess multi-stakeholder collaborations.  
\item An MBSE-HFGT workflow can engage communities and institutions in participatory (graphical) modeling and support collaboration through mathematical analysis (as evidenced by the recently funded NSF GCR project).  
\item HFGT supports a wide range of continuous and discretized supply chain models; offering a comprehensive framework that integrates these critical aspects in megaproject engineering management into a unified system for analysis and optimization.
\end{itemize}
By addressing Denicol et. al's concluding recommendation, MBSE and HFGT not only align with current megaproject management research needs but also push the boundaries of how megaprojects can be managed and optimized within a unified theoretical foundation.  

\bibliographystyle{IEEEtran}
\bibliography{LIINESLibrary, LIINESPublications-2,PM-references}
\end{document}